# Motion rejection and spectral unmixing for accurate estimation of *in vivo* oxygen saturation using multispectral optoacoustic tomography.

Mitradeep Sarkar, Mailyn Pérez-Liva, Gilles Renault, Bertrand Tavitian and Jérôme Gateau

*Abstract*— **Multispectral Optoacoustic Tomography (MSOT) uniquely enables spatial mapping in high resolution of oxygen saturation ($SO_2$), with potential applications in studying pathological complications and therapy efficacy. MSOT offers seamless integration with ultrasonography, by using a common ultrasound detector array. However, MSOT relies on multiple successive acquisitions of optoacoustic (OA) images at different optical wavelengths and the low frame rate of OA imaging makes the MSOT acquisition sensitive to body/respiratory motion. Moreover, estimation of $SO_2$ is highly sensitive to noise, and artefacts related to the respiratory motion of the animal were identified as the primary source of noise in MSOT.**

**In this work, we propose a two-step image processing method for $SO_2$ estimation in deep tissues. First, to mitigate motion artefacts, we propose a method of selection of OA images acquired only during the respiratory pause of the animal, using ultrafast ultrasound images (USIs) acquired immediately after each OA acquisition (USI acquisition duration of 1.4 ms and a total delay of 7 ms). We show that gating is more effective using USIs than OA images at different optical wavelengths.**

**Secondly, we propose a novel method which can estimate directly the $SO_2$ value of a pixel and at the same time evaluate the amount of noise present in that pixel. Hence, the method can efficiently eliminate the pixels dominated by noise from the final $SO_2$ map.**

**Our post-processing method is shown to outperform conventional methods for $SO_2$ estimation, and the method was validated by *in vivo* oxygen challenge experiments.**

M.Sarkar, M.Pérez-Liva and B.Tavitian are with **Université Paris Cité, PARCC, Inserm U970, F-75015 Paris, France** . M.Sarkar is now with ICFO-Institut de Ciencies Fotoniques, Castelldefels (Barcelona), Spain. M.Perez-Liva is now with the Nuclear Physics Group and IPARCOS, Department of Structure of Matter, Thermal Physics and Electronics, CEI Moncloa, Universidad Complutense de Madrid, 28040 Madrid, Spain.
B.Tavitian is also with the Radiology Department, AP-HP, Hôpital européen Georges Pompidou, F-75015 Paris.
   G. Renault is with the Institut Cochin, Université Paris Cité, INSERM, CNRS, 75014 Paris, France.
   J. Gateau is with the Laboratoire d'Imagerie Biomédicale, Sorbonne Université, CNRS, INSERM, LIB, 75006 Paris, France (e-mail: jerome.gateau@sorbonne-universite.fr).

## I. INTRODUCTION

Multispectral Opto-acoustic tomography (MSOT), which relies on the photoacoustic effect and the spectral response of chromophores, offers the possibility of high resolution mapping of *in vivo* oxygen saturation ($SO_2$) in a non-invasive manner and without the use of external contrast enhancing agents [1]. MSOT is performed with a sequential excitation of the imaged region with laser pulses at different optical wavelengths. The ultrasound (US) field generated by each laser pulse is recorded to reconstruct an optoacoustic (OA) image. In a second step, the spectral unmixing of the OA images enables to retrieve the abundance of the different optically absorbing compounds in the tissue. On the other hand, US imaging (USI), based on the backscattering of US waves, enable the acquisition of information complementary to MSOT on deep tissues *in vivo* such as the anatomical context provided by the echogenicity of different tissue structures [2][3]. Using a linear US array, dual mode US / OA imaging systems have been developed [4] by several research groups for clinical [5] [6] and pre-clinical applications[7].

US imaging can be performed at ultrafast acquisition rates of several thousand frames per second [8]. In contrast, the frame rate in OA imaging is limited by the laser repetition rate, commonly. on the order of 10 to 50 pulses per second. As a consequence, significant motion can occur during the successive OA acquisitions of a MSOT sequence, especially in small animal imaging with a high breathing rate under anesthesia [9]). In MSOT, spectral variations are measured for each pixel. Therefore, motion between the multiple laser shots is expected to cause severe degradation of the final image quality [10].

Tracking algorithms are commonly used for motion correction, applied either on the OA spectral images or on interleaved USI [10][11]. However, such algorithms are complex and may result in erroneous motion correction in particular when tissue displacements are perpendicular to the imaging plane. Self-gating of OA images (at a single wavelength) has been reported previously [12], but the method requires a consistent spatial distribution of pixel values between the selected images. Due to the different absorption spectra and different spatial distributions of the two main endogenous chromophores, i.e., deoxyhemoglobin (Hb) and oxy-hemoglobin ($HbO_2$), images acquired successively at different optical wavelengths are not perfectly correlated compared to images acquired at the same wavelength. Therefore, self-gating of OA images cannot be directly applied to MSOT when multiple optical wavelengths are involved.

To mitigate the motion artifacts in MSOT, we propose here to use a bimodal acquisition combining MSOT and plane-wave ultrafast USI acquired immediately after each OA acquisition. Plane-wave acquisition allows to provide a snap shot of the entire imaged area only a few milliseconds after the OA acquisition. We performed a frame selection on the USIs, using a self-gating approach to keep only the frames acquired during the respiratory pause of the animal. Finally, we applied the selection on the OA images to obtain a MSOT dataset for the evaluation of the oxygen saturation.

The oxygen saturation ($SO_2$) is defined as the ratio of the $HbO_2$ concentration over the total hemoglobin concentration ($HbO_2$ and $Hb$). Conventional methods for the $SO_2$ evaluation with MSOT datasets determine the absolute spatial abundances of $Hb$ and $HbO_2$ before computing the ratio [13]. Currently, the most commonly used method is the least square method that determines separately the concentration of $Hb$ and $HbO_2$ from the MSOT images on a pixel per pixel basis using an unmixing algorithm [14]. The spectrum at a given pixel in the MSOT image is decomposed in the linear sum of the known spectra $S_k(\lambda)$ of the different absorbing components with weights corresponding to the molar concentration $C_k$. The weights are determined by minimizing the sum of the squares of the residuals with a non-negative constrain (non-negative least squares). This least square method does not offer any reliable parameter to eliminate pixels in which data artefacts corrupt the estimated concentration. The presence of such erroneous pixels compromises the identification of the vasculature or other objects of interest in the final $SO_2$ image, [15]. Usually, an amplitude thresholding is used to eliminate such pixels but it appears inadequate in most cases.

Here, we propose a method to estimate directly the $SO_2$, based on the assumption that hemoglobin is the dominant absorber in the imaged region and that spectral noise can be modelled as an additive flat spectrum. Our method is based on quantifying the amount of spectral deviation present in each pixel for different expected values of $SO_2$. The value corresponding to the smallest deviation allows us to identify the actual $SO_2$ value of the pixel. Furthermore, a threshold at the acceptable spectral deviation eliminates those pixels which are dominated by noise from the final $SO_2$ image.

Using these two new image processing methods, i.e., gating multispectral OA images to remove motion related artifacts, followed by pixel selection, we show using experimental data recorded *in vivo* in small animals that our $SO_2$ estimations are more robust and accurate than conventional methods.

## II. METHODS

### A. Animal handling and experiments.

Mouse experiments were approved by the ethics committee for animal experimentation of Université Paris Cité and registered by the French Ministry for Higher Education and Research under reference number APAFIS#31847-2021040118582067. All animals were housed in a conventional animal facility under a 12h light and 12h dark cycle and controlled temperature (22 °C). Animals were fed and watered *ad libitum*.

All imaging was done in a non-invasive manner, with the mouse anesthetized using 2% Isoflurane and medical air mixture (20% $O_2$). During the experiments the mice were positioned on a thermalized platform and the body temperature was assured to be at 37°C by monitoring with a rectal probe. For $O_2$ challenge experiments, the mice were made to breathe 100% $O_2$ (hyperoxia), medical air mixture (normoxia) or a mixture of 7% O2 and 93% N2 gas (hypoxia). For hyperoxia challenge, the mice were stabilized for 5 minutes before image acquisitions, while the hypoxia challenge was induced for less than 4 minutes, and it was verified that the animals recovered after the challenge. For the experiment post-mortem (1 experiment), the anesthetized mouse was positioned on the imaging platform and then sacrificed with an overdose of $CO_2$. The MSOT/USI sequence was conducted immediately after the euthanasia.

### B. Imaging system and acquisition sequence.
#### 1) Imaging system.

A clinical linear US transducer array with a center frequency of 7.8 MHz (L12-5, 50mm, ATL) was chosen to have acceptable performance for both MSOT and pulse-echo USI in mice. The 128 central elements (out of 256) of the multiplexed transducer array were used to generate 2D cross-sectional (transverse) images over an image width of 25 mm. The transducer was driven by a programmable US system (Vantage 256, Verasonics). The imaging depth range was set between 5 and 16 mm.

Optical excitation for the OA imaging was performed with an OPO tunable nano-second pulsed laser (680nm-980 nm, pulse repetition frequency (PRF): 20Hz, Innolas). The laser light was guided towards the imaged region using a bifurcated bundle of optical fibers. Each leg of the bundle was held with a home-built mount comprised of a fiber housing and a mirror to obtain a rectangular spot of $20\times11mm^2$ (h×w) perpendicular to the fiber's directions (Fig 1a). The output fluence of each end was about 8 mJ/cm$^2$ which is below the Maximum Permissible Exposure (MPE) limits for the skin given the laser PRF. For the MSOT acquisitions, the pockel cell of the laser (Q-switch) was externally triggered by the US system, which precisely synchronize the US acquisition [16]. OA signals were recorded with a sampling rate of 62.5 MS/s, and a constant gain (no time gain compensation). The per-pulse energy was recorded using an uncalibrated (but linear) internal energy monitor and used to correct the OA signals for the per-pulse laser energy fluctuations.

A compound plane-wave USI was acquired 5 ms after every laser pulse using seven tilted plane waves with angles equally spaced between -3.15° to +3.15°. A delay of 200 µs between each plane wave acquisition resulted in a total duration of the acquisition of one frame of 1.4 ms. The global time delay between the OA and the US acquisition was then around 7 ms. US excitation corresponded to a 1 half-cycle at instrument with the ultrasound array positioned on the dorsal side of the mouse. b) The acquisition sequence where each OA acquisition (red arrows) is followed by a plane-wave US acquisition (blue line). For OA acquisition, 18 optical wavelengths (sweep) was repeated 7 times. c) An example of the image stacks obtained during an experimental session. Each OA image (acquired for each laser pulse) has a corresponding USI. These US frames were correlated among themselves to reject OA images with motion. The OA images selected after motion rejection, for each wavelength, were averaged to obtain the stacks of multispectral (MSOT). Spectral unmixing was done on these images by the variance method to generate the final $SO_2$ image. The absorption spectra for the molecules expected to be present in a mouse are also shown. For Hb and $HbO_2$, 150g Hb/l (whole blood) is considered.

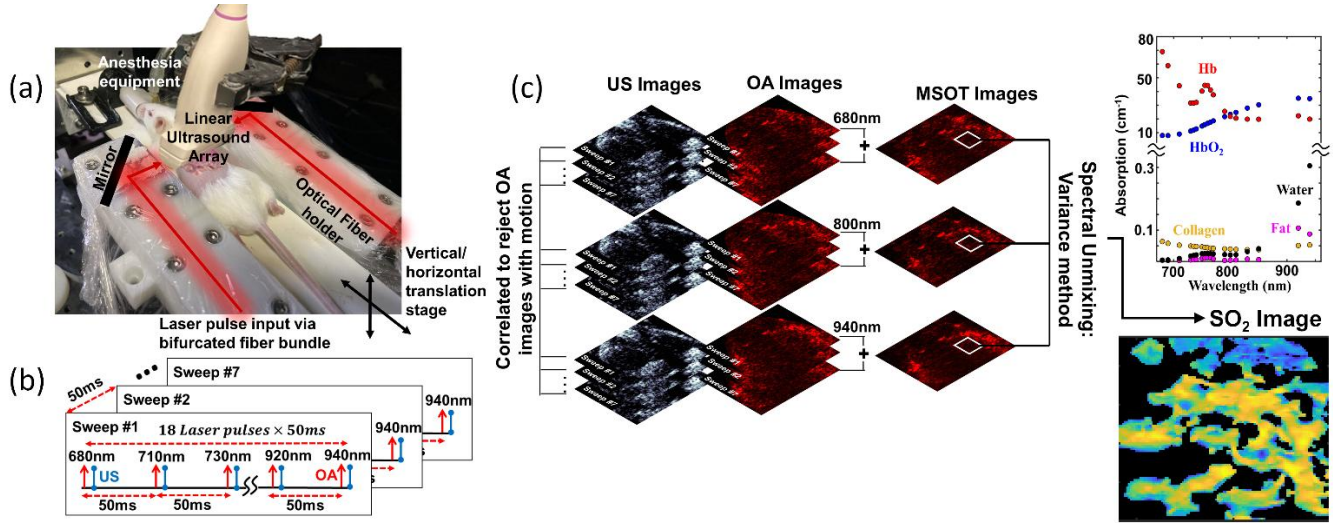

7.8MHz. US signals were recorded at 31.2 MS/s. The time gain compensation (TGC) was gradually increased for depth between 5 mm and 8 mm and set constant from 8 mm.

The skin of the shaved mouse was illuminated along the full body transverse section. We imaged the kidney regions from the dorsal side. The illumination system was independent from the ultrasound array to facilitate the application of ultrasound gel between the skin and the array for acoustic coupling. The bundle ends were aligned to illuminate the imaging plane and were positioned on the sides of the array (Fig 1a).

*2) Image acquisition.*

The acquisition sequence is shown in Fig 1b. 18 optical wavelengths were used: 680, 690, 710, 730, 735, 740, 750, 755, 760, 765, 770, 790, 800, 810, 830, 850, 920, 940 nm, with finer sampling close to the inflection (760nm) of the deoxyhemoglobin spectrum (Fig. 1c). The MSOT acquisitions corresponds to 7 consecutive sweeps of the 18 laser pulses, for a total of 126 laser pulses and an acquisition duration of 6.3 s. For each sweep, consecutive pulses corresponded to different optical wavelengths.

*3) Image reconstruction*

For both OA and US imaging, envelope detected images were reconstructed using conventional two-dimensional delay and sum (DAS) beamforming algorithms with a pixel size of 66μm×66μm. A lateral dynamic aperture (f-number = 0.5) was applied. OA signals were band-pass filtered between 2MHz and 15MHz (3rd order, Butterworth) for noise removal.

For the USI, the tilted plane wave acquisitions were separately beamformed and then coherently summed to obtain the final USI[17]. The DAS algorithms for US and OA data used the same speed of sound, but different time offsets to account for delays linked to the lens of the US array and to a synchronization delay between the laser firing and the US acquisition. The time delays ensured superimposition of the US and OA images for an object with a dual contrast.

The OA signals were corrected prior to the image reconstruction for the per-pulse laser energy fluctuations of the laser. Additionally, we determined experimentally wavelength-dependent correction coefficients to account for the mean energy variations of the laser output with the wavelength and the spectral responses of the energy monitor and the multimode fibers. For this purpose, we imaged India ink diluted in water and injected in a PTFE tube phantom. The 7 OA images acquired at the same wavelength were incoherently averaged. We obtained the correction coefficients at each wavelength by computing the ratio between the known spectrum of India ink[18] and the measured amplitude of the tube on the images. These determined correction coefficients were then used for all the datasets.

*C. Image post-processing*

*1) Selection of frames based on the Pearson's coefficient*

A motion rejection algorithm based on the Pearson's linear correlation coefficient computed on the USIs was implemented. The Pearson's linear correlation coefficient is commonly used to detect and quantify disparity between two images[19]. The coefficient for two images i and j, each with N pixels is given as

$$R_{i,j} = \frac{\sum_1^N (i_n - \rho_i)(j_n - \rho_j)}{\sqrt{\sum_1^N (i_n - \rho_i)^2 \sum_1^N (j_n - \rho_j)^2}} = \frac{cov(i,j)}{\delta_i \delta_j} \quad (1)$$

Where ρ and δ are the mean and standard deviations respectively, $R_{i,j}=1$ denotes that the two images are spatially superimposed and $R_{i,j}$ decreases with the increasing spatial decorrelation. The 126 envelope-detected USIs of an image sequence were arranged in a 2D matrix $S_{US}$ and $R_{i,j}$ was calculated for all pairs of (i,j) by the autocorrelation of $S_{US}$. The results were arranged in a 126×126 matrix ($R_{US}$).

Then, we determined which image will serve as a reference for the frame selection. The frame selection aims at selecting the maximum of spatially correlated MSOT images to: 1) average incoherently OA images acquired at the same optical wavelength to increase the image signal-to-noise ratio, 2) obtain spatially correlated images at the 18 optical wavelengths and then compute the map of $SO_2$.

For each row k of the $R_{US}$ matrix, we assigned $R_k$ as the highest $R_{k,j}$ value with $k \neq j$. Then among all $R_k$, we evaluated the highest value for which at least 3 out of 7 sweeps would be selected for all the 18 wavelengths (Fig 2c). The row k with that highest $R_k$ value was selected as reference, and the value was defined as a threshold $R^{th}$.

Frames with $R_{k,j} > R^{th}$ were assigned to the static group and the others to the motion group. Frame incoherent averaging was performed in the static group and the motion group was discarded. The minimum $R^{th}$ obtained for all experiments presented in this paper was 0.87.

It worth noting that the Pearson's linear correlation coefficients were computed on the USIs while the selection of frame is subsequently applied to OA images. The global time delay of 7 ms between the OA and plane wave US acquisitions is short enough to assure negligible movement between the two images and at the same time, long enough to prevent ultrasound interference between the two modalities. Healthy mice under anesthesia have respiratory rates of 80-120 breaths per minute, thus 1 respiratory cycle lasts 500ms-750ms. The OA and USIs acquired with a time delay of 7 ms can then be spatially superimposed.

*2) Selection of pixels for $SO_2$ evaluation, by variance method*

$SO_2$ was evaluated from the OA images belonging only to the static group (selected) after averaging of the images acquired at the same wavelength. The $SO_2$ is defined as the ratio between the local concentration of $HbO_2$ ($C_{HbO_2}$) and the local concentration of total hemoglobin, both oxygenated and deoxygenated ($C_{Hb}$). Therefore:

$$SO_2 = C_{HbO_2} / (C_{Hb} + C_{HbO_2}) \quad (2)$$

We developed a method to exclude pixels for which the spectrum is dominated by noise. Our main assumption is that noise can be modeled by the addition of a flat spectrum (constant values for all the wavelength) with a mean value over the wavelength equal to zero. We also choose to

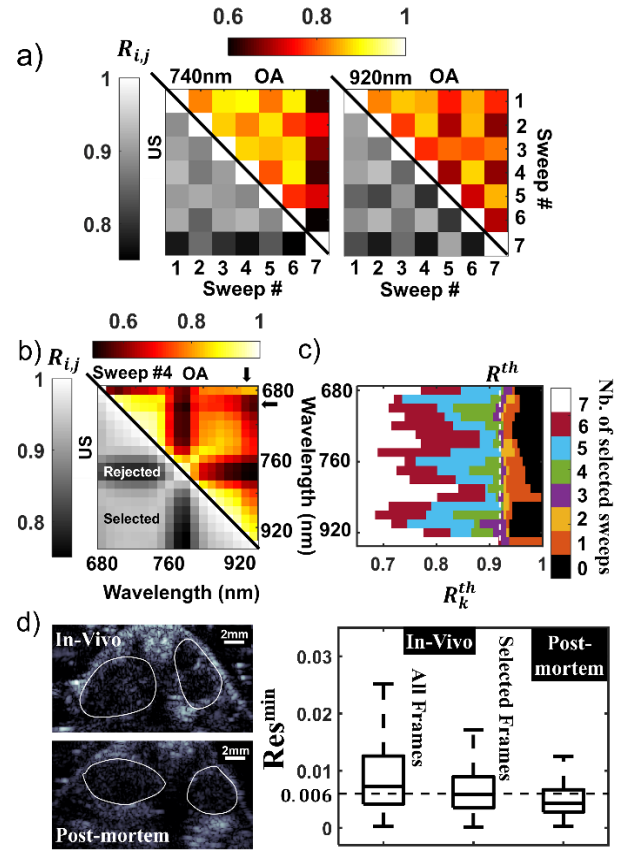

Fig 2. a) The correlation coefficients ($R_{i,j}$) for pairs of OA images (top half, color) acquired at a wavelength λ (740nm, left and 920nm, right) during the 7 sweeps. The $R_{i,j}$ for the corresponding pairs of USIs is also shown (bottom half, gray scale). b) Correlation coefficients for pairs of OA images and USIs acquired during one sweep. The black arrows indicate pairs of images where decorrelation occurs only for the OA modality. c) The number of images (sweeps) per wavelength with $R_k > R_k^{th}$. The overall threshold $R^{th}$ is shown by dashed white line. d) $Res^{min}$ values for an in vivo experiment, with and without motion rejection. The mouse kidney was selected by manual segmentation of the B-mode (US) image (shown in left, and the ROI are shown by white lines, number of pixels in ROI was 17825, the respiratory motion rejection thresholds ($R_{i,j}^{thresh}$) was 0.92). $Res^{min}$ values for the experiment post-mortem (same as Fig 3, for 15516 pixels in ROI) is also shown. The threshold at Res=6×10⁻³ is shown by the black dashed line. One can note that the cross-sectional images of mice does not guarantee the symmetrical positioning of the two kidneys in determine directly the $SO_2$ value to have a single parameter problem.

Mice used for this work were of albino strain and lipid [20], collagen [21] and water [22] have low absorbance with featureless spectra (almost flat spectra) in the wavelength range used, hence spectra are dominated by hemoglobin (Fig. 1). Then, the measured MSOT spectrum $P^{obs}_m(\lambda)$ for pixel m, can be written, following the linear model as

$$P^{obs}_m(\lambda) = A_m \cdot [P^{model}(\lambda, SO_2) + \psi_{0,m}(\lambda)] \quad (3)$$

with $P^{model} = (1 - SO_2) \cdot S_{Hb}(\lambda) + SO_2 \cdot S_{HbO_2}(\lambda) \quad (4)$

Where $S_{Hb}(\lambda)$ and $S_{HbO_2}(\lambda)$ are the spectra of Hb and $HbO_2$ respectively. $A_m$ is a wavelength independent factor which includes the incident light fluence, the photothermal conversion efficiency, local concentrations of hemoglobin and the Grüneisen coefficient. $\psi_0(\lambda)$ encompasses all parasitic spectral deviations (electronic noise, noise induced by slight displacement between the images at different wavelengths, presence of other absorbers) and is assumed to have a zero-mean value with respect to λ. Because the value $A_m$ is complex to determine, we normalized the spectra $P^{obs}_m(\lambda)$ and $P^{model}(\lambda)$ by their respective mean value with respect to λ. We obtain $\overline{P^{obs}(\lambda)}$ and $\overline{P^{model}(\lambda)}$, and define $D_m$ as their absolute difference:

$$D_m(\lambda, SO_2) = \left|\overline{P^{obs}(\lambda)} - \overline{P^{model}}(\lambda, SO_2)\right| \quad (5)$$

$D_m$ is proportional to $|\psi_0(\lambda)|$ for the correct $SO_2$. The variance ($var$) of $D_m$ is called the Residual (Res) and is given by:

$$Res_m(SO_2) = \frac{1}{n_\lambda} \sum_{i=1}^{n_\lambda} \left|D_m(\lambda_i, SO_2) - \mu_{D_m}(SO_2)\right|^2 \quad (6)$$

Where $\mu_{Dm}$ is the mean value of $D_m$ over the wavelengths.

For each pixel m, Res values were evaluated for $SO_2$ between 0 and 100% with steps of 1%, and the final $SO_2$ is estimated by the value for which Res is minimum ($Res^{min}$). The method was called the Variance method. The benefit of this method is that as only one component ($SO_2$) is considered, the minimization problem leads to a unique solution. The drawback is that unlike the least square method, our method does not calculate the absolute spatial abundances of Hb and $HbO_2$.

We found that a threshold can be applied to Res to exclude pixels corrupted by noise. Owing to the uniqueness of the solution for the variance method, the absolute value of $Res^{min}$ indicates the accuracy of the overall $SO_2$ estimation. Smaller the value of Res, better is the estimation of $SO_2$.

The spectrum for $SO_2$=60% has the least variance and hence for a flat spectrum that correspond to $\overline{P^{obs}(\lambda)} = 1$, the solution is $SO_2$=60% and the residual: $Res^{min}=var(|1 - \overline{P^{model}}(\lambda, SO_2 = 60\%)|) = 6\times10^{-3}$ for the 18 wavelengths used. Any observed spectra which results from absorbers other than blood (Hb or $HbO_2$ and their linear combinations) or corrupted by spectral coloration will have a $Res^{min} > 6\times10^{-3}$.

To validate the number of optical wavelengths, a PTFE tube (300μm inner diameter) filled with India ink was imaged and the corresponding $Res^{min}$ for the image was calculated. It was seen that the median value of $Res^{min}$ equals $6\times10^{-3}$ at the position of the tube for $SO_2$=60% and $n_\lambda$=18 (see Appendix B).

Considering $\psi(\lambda) = \overline{P^{obs}(\lambda)} - \overline{P^{model}}(\lambda, SO_2)$, which is proportional to $\psi_0(\lambda)$, to be a normal distribution (Appendix C) with a mean μ=0 and a standard deviation σ, $Res^{min} = 0.36\sigma^2$ (see Appendix A). Hence, thresholding with $Res^{min} \leq 6\times10^{-3}$ will select all pixels for which the relative standard deviation ($100 \times \sigma/\mu(\overline{P^{obs}})$ %) of $\psi(\lambda)$ is lower than 13%.

For the Variance method, a binary mask for reliable pixels for the $SO_2$ estimation was defined by putting all the pixels with $Res^{min} > 6\times10^{-3}$ equal to zero. For comparison, the least square method was applied and a binary mask was also defined by setting to zero, pixels for which the sum of the abundances $C_{Hb} + C_{HbO_2}$ was below 0.1 of the normalized values (normalization by the maximum over the ROI). Then, a gaussian filter using Otsu's method with global threshold of 0.5, coupled with area opening algorithm to remove unconnected objects less than 30 pixels was used to smoothen the binary masks.

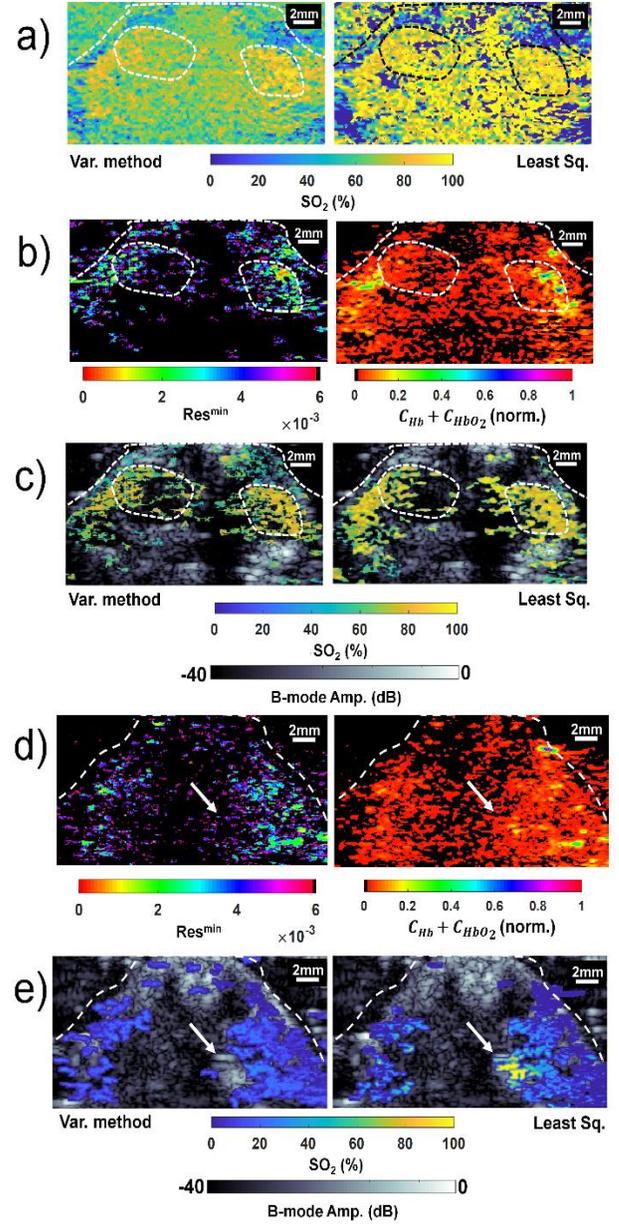

Fig 3. a) $SO_2$ estimations using the variance method before thresholding (left) and non-negative constrained least square inversion (right) for mouse breathing air. b) The image of $Res^{min}$ values evaluated using the variance method (left) and the normalized sum of the abundances of Hb and $HbO_2$ obtained by the least square method (right). c) The SO2 map obtained by the two respective methods superposed on the anatomical B-mode image. The contour of the kidneys obtained by manual segmentation of the anatomical image is shown by white dashed lines. d) Same as b for postmortem. e) The SO2 maps by the two methods postmortem. The arrow indicates the region where the least square method fails in thresholding ambiguous pixels. The experiments *in vivo* and postmortem was done on different mice. For all the images, the surface of the mouse is delineated. The image widths are 25mm and the top of the image is at 5mm while the bottom is at 16mm from the ultrasound transducer. Fig 1. a) Annotated picture of the bimodal

*3) Simulation to verify the accuracy of the variance method*

Since having calibrated $SO_2$ phantoms is experimentally challenging, we performed a numerical simulation to assess the accuracy and the precision of the variance method. For a given value of SO₂, a multispectral image (512×512 pixels, 18 wavelengths) was created using Eq.4: $\overline{P^{model}}(\lambda, SO_2)$ was computed for the 18 wavelengths and it was then multiplied by an amplitude mask (512×512 pixels) to mimic the heterogeneous amplitude distribution in the images. To obtain an amplitude mask with spatial features similar to the MSOT images, we generated an intensity map of a Gaussian random wavefield (speckle pattern) with a speckle grain size of 71 pixels. The mask was obtained by computing the two-dimensional Fourier transform of a circular aperture addressed with random phases, and the intensity map was normalized by its maximum value. Randomly generated noise $\psi(\lambda)$, having a normal distribution with a mean μ=0 and a standard deviation σ=0.12, was added to each pixel of the image and at each wavelength. The variance and the least square methods were applied to the generated image. For the Variance method, all the pixels with $Res^{min} > 6\times10^{-3}$ were rejected. For the least square method, pixels for which the sum of the abundances $C_{Hb} + C_{HbO_2}$ was below 0.1 of the maximum value in the image were rejected.

## III. RESULTS

### A. Respiratory motion rejection for MSOT using interleaved compound UUS frames.

The correlation coefficients $R_{i,j}$ between the 7 OA images acquired at the same optical wavelength (λ) but for different sweeps are shown in Fig 2a along with the correlation coefficients of the respective USIs. The pairs of OA image which are highly decorrelated (drops in the value of $R_{i,j}$) correspond perfectly with the decorrelated pairs of USIs which confirms that the USIs can be used to reject OA images acquired during the respiratory motion.

Using USIs has several advantages. First, the magnitude of $R_{i,j}$ for decorrelated images was found to be wavelength dependent for the OA images whereas it is wavelength independent for the USIs (Fig 2b). This results in a greater number of decorrelated OA images than USIs. This is due to the fact that while the decorrelation of the USIs is purely related to the respiratory motion of the animal, that of the OA image also include the inter-wavelength decorrelation of the absorption spectra of Hb and HbO₂. Hence, using USIs to identify instances of respiratory motion of the animal is much more robust and reliable than applying motion rejection based on the OA images.

In Fig 2b, we show the sorting of the acquired images into two groups namely a static (selected) group and a motion (rejected) group. The images of the static group are considered to be acquired during the respiratory pause of the animal and only those OA images are selected for the spectral unmixing process. A threshold ($R^{th}$) was defined to separate the two groups. The static group is defined by $R_{i,j} > R^{th}$. Because of the variability between experiments, the value of $R^{th}$ was not fixed but chosen with the criterion of inclusion of at least 3 images (sweeps) per optical wavelength in the static group (Fig 2c).

Furthermore, using the temporal variation of $R_{i,j}$ values of a chronologically ordered set of USI pairs, the respiratory cycle duration can be derived as the time difference between consecutive $R_{i,j}$ minima. This helps to monitor the respiratory rate of the animal and serves as an additional pathological data of the animal.

### B. SO₂ estimation using the Variance Method.

For *in vivo* OA images, the pixels inside the kidneys were selected with a manual segmentation on the USIs (Fig 2d). The $Res^{min}$ calculated in the selected area is shown in Fig 2d. The median of $Res^{min}$ values without motion rejection is 7.3×10⁻³ while, with motion rejection, the median of $Res^{min}$ was evaluated to be 5.7×10⁻³.

An experiment was performed 1 minute after a mouse was

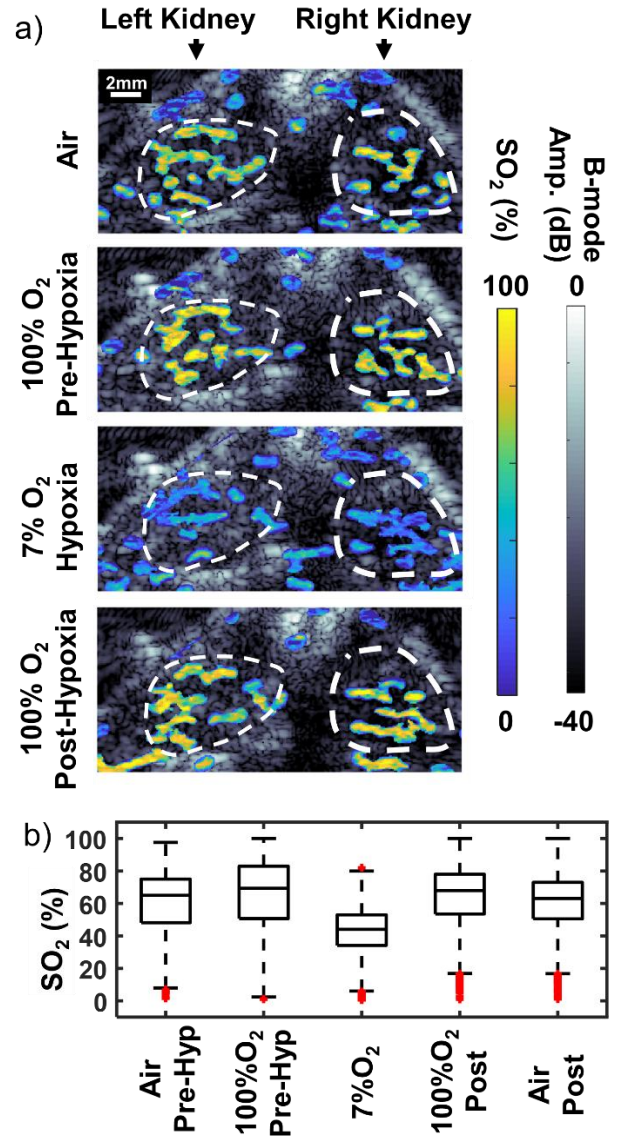

Fig 4. a) Overlay of SO2 and B-mode images of the two kidneys during normoxia (breathing air), hypoxia (7% O₂, 93% N₂ gas) and hyperoxia (100% O₂) induction. The results for SO2 obtained by the variance method are shown. b) An ROI encompassing both the kidneys was manually segmented using the corresponding B-mode image in normoxic conditions (white dashed lines). The statistics of the number of pixels of the fixed ROI for SO2 values during the O₂ challenge is represented by the boxplot. The total number of pixels in the ROI was 13492. The percentage of outliers is below 2% of the euthanized (post mortem) and the median of the Res$^{min}$ values was found to be 4.2×10$^{-3}$. We verified by calculating $\psi(\lambda)$ with $\overline{P^{model}}(\lambda, SO_2 = 0)$ that the relative standard deviation equals 10.7% and thereby corresponds to the theoretical value (11%) obtained with a normal distribution (Appendix C). For the *in vivo* experiments, the relative standard deviation on $\psi(\lambda)$ was verified for $\overline{P^{model}}(\lambda, SO_2)$ using the $SO_2$ values evaluated by the variance method and it was found equal 14.2% before motion rejection and 12.5% after motion rejection. Hence, thresholding with Res$^{min}$ =6×10$^{-3}$ is expected to be efficient for $SO_2$ evaluation after motion rejection.

The basic challenge encountered by linear unmixing models for *in vivo* imaging and the effectiveness of the variance method in overcoming them is demonstrated in Fig 3. As mentioned earlier, for the variance method, flat spectra or white noise dominated spectra (with zero mean deviations) lead to an estimated $SO_2$ value of 60%. This is evident in Fig 3a since the presence of $SO_2$=60% is almost homogenously throughout the $SO_2$ map even in the parts of the image with only ultrasound gel.

For the least square method, we used a linear unmixing algorithm with a non-negative constrain. A thresholding was done on the sum of the abundances ($C_{Hb} + C_{HbO_2}$) at 0.1 of the normalized values (Fig 3b). The value of threshold was chosen empirically to have a compromised balance between accepted and rejected pixels in the kidneys, however such thresholding is far from efficient and also not mathematically robust. By comparison, a thresholding at Res$^{min}$ = 6×10$^{-3}$ effectively eliminates ambiguous $SO_2$ estimations which furthermore validates the variance method. Even though the result obtained using the least square method may seem acceptable for Fig 3c, the variance method is imperative mainly for $SO_2$ estimation accuracy in hypoxic conditions. In Fig 3d-e we show that the conventional least square method overestimates the global $SO_2$ for anoxic post mortem conditions.

### C. Validating $SO_2$ estimation with numerical simulation

The accuracy and precision of the variance method were assessed by numerical simulation to obtain the ground truth of the $SO_2$ value. Five different $SO_2$ values were simulated to cover the entire range from 0 to 100%. After pixel rejection, the mean and standard deviation of the $SO_2$ values for the selected pixels were calculated for the variance method and the least square method, respectively. Results are shown in Table.1. For the variance method around 10% of the pixels were selected for all the 5 simulated $SO_2$ values while > 65% of pixels were selected for the least square method, with our pixel rejection methods. The lower mean error and lower standard deviations of the variance method validate our claims regarding the higher accuracy and the higher precision, respectively, compared to the conventional unmixing method. These performances are mainly due to the pixel rejection method.

Table 1: Mean error (SO$_2$ {Estimated}-SO$_2$ {Ground Truth}) and standard deviation of estimated SO$_2$ values using the variance method and the least square method on the numerically simulated multispectral images.

| Ground Truth $SO_2$ | Mean error (%) | | Standard deviation (%) | |
|---|---|---|---|---|
| | Least Sq. | Var. | Least Sq. | Var |
| 0% | 21 | 5 | 26 | 6 |
| 25% | 9 | 0 | 21 | 8 |
| 50% | 2 | 0 | 15 | 6 |
| 75% | -1 | 0 | 10 | 5 |
| 100% | -4 | -2 | 7 | 2 |

### D. Validating $SO_2$ estimation with $O_2$ challenge experiments..

To further validate the accuracy of the oxygen saturation measurements and to observe the spatial deoxygenation of the renal blood vessels, hypoxia (mouse breathing 7% oxygen, 93% nitrogen gas) and hyperoxia (100% oxygen) challenges were conducted on the same mouse consecutively. We performed the same experimental protocol in three different mice. The images for one representative mouse are shown in Fig.4. The motion rejection and variance method for unmixing was applied to the entire image. The statistical distribution of the $SO_2$ estimate for this mouse in the ROI selected by manual segmentation (white dashed lines in Fig 4a) is presented with boxplots in Fig 4b. One can note that the illumination was preferentially directed towards the kidney region (marked by white dashed lines in Fig 4a). So, the upper part of the image which was illuminated by the edge of the beam may result in erroneous estimation of $SO_2$. The large spread of the $SO_2$ values in the ROI can be attributed to the fact that the vasculature of the kidneys cannot be resolved so we measure the overall $SO_2$ which includes both the arteries and veins.

Global statistical evaluations of the $SO_2$ estimates were conducted on the images acquired for the three mice. For each mouse, the pixels were selected for both kidneys (by manual segmentation of B-mode image), leading to a total of 3×2=6 ROIs. For each ROI, the median, upper and lower quartiles of $SO_2$ values were evaluated. For each quantifier, we computed the mean and the coefficient of variation (COV, standard deviation over mean) over the 6 values.

In normoxic conditions, we obtained a median $SO_2^m$=64.6% (COV = 2.5%), upper quartile $SO_2^u$=74.3% (COV=2.9%) and lower quartile $SO_2^l$=50% (COV=3%). The mean value of $SO_2$ in normoxic conditions, measured by OA was reported elsewhere to be 64% for healthy mice [23]. This is consistent with our estimation of $SO_2$ (Fig. 4).

For hyperoxia, we obtained a $SO_2^m$=69% (COV=4%), upper quartile $SO_2^u$=83% (COV=5%) and lower quartile $SO_2^l$=52% (COV=5%). In the literature, the peripheral $SO_2$ for healthy mice is reported to change in the range from 72% to 80% in normoxic and hyperoxic conditions [24]. This is consistent with the $SO_2^u$ value obtained by the Variance method.

Under hypoxic conditions, the mean $SO_2$ after 200 seconds of hypoxia ($SO_2^m$) was 44% (COV=4.7%), $SO_2^u$=55% (COV=5%) and $SO_2^l$=30% (COV=5%). These values are very close to the ones found in the mouse mesometrium (uterus) [25]. The COV for normoxic conditions is lower than during the $O_2$ challenge experiment, and this is expected as the adaptations of different animals to breathing challenges may be different.

## IV. DISCUSSIONS

In this work, we report a novel method of respiratory motion correction for MSOT, which relies on interleaved USIs. Respiratory motion causes a tissue displacement of more than 300μm (estimated during our experiments from the displacements between selected and rejected frames) which results in inaccurate estimations of $SO_2$. To avoid distortion of the MSOT images, the respiratory pause of the animal was identified using the interleaved USIs, and only the MSOT images acquired during the respiratory pause of the animal were retained for the spectral unmixing. A threshold for motion rejection was defined as a trade-off between the maximum correlation among the selected images and selecting at least 3 images per excitation optical wavelength.

The Pearson's linear correlation coefficient ($R_{i,j}$) was used to quantify disparity between the images, and we found that the threshold $R^{th}$ for all experiments ranged from 0.87 to 0.94. This is due to respiratory rate variations between mice and experimental conditions. An average correlation of 0.92 was observed for all experiments during the respiratory pause of the animal. This decorrelation can be attributed to the subtle respiratory motion during the respiratory pause of the animal.

The use of interleaved US images to account for tissue motion in multispectral optoacoustic imaging has been recently proposed [10] for two optical wavelengths. In that study, regional motion correction was evaluated on the US images and subsequently applied on the OA images. The method was found operational for small displacement (imaging a human arm). For the larger respiratory motion of a mouse and the large number of optical wavelengths used in our study, self-gating was determined to be easier to implement. Regional motion correction will be envisioned in future studies to further refine the motion correction for images with a Pearson's linear correlation coefficient close to the threshold.

Numerous spectral unmixing method can be applied to the MSOT images to compute the $SO_2$ [13], and most have similar accuracy values [26]. The accuracy of $SO_2$ estimations for pixels with high SNR is similar for most available unmixing methods. However, for robust MSOT imaging, it is extremely important to be able to discriminate pixels with reliable $SO_2$ estimations from noise dominated pixels. We presented a novel method of spectral unmixing dedicated for the $SO_2$ evaluation. A linear unmixing model could be applied here with the assumption that spectral coloration can be neglected for preclinical imaging and for the range of wavelengths used [27].

The method assumes that the dominating absorber is hemoglobin and that the spectral noise can be considered to be additive with similar statistical properties at every wavelength. The variance method calculates the variance of the difference ($D_m(SO_2)$) between the measured spectra and a set of model spectra with $SO_2$ values between 0 and 100%. The minimum of the variance gives the estimated $SO_2$ value while its absolute value ($Res^{min}$) represents the amount of spectral deviations present in the measurement. The spectral deviations present in our MSOT images was measured to have a relative standard deviation (RSD) of 10.7% for an experiment conducted post-mortem. This spectral deviation is quite acceptable and publications elsewhere on MSOT show similar or higher noise level for unfiltered images [28][29].

The respiratory motion of the animal was identified as a major source of spectral deviations. Without motion rejection, the RSD was found to be higher than that with motion (Fig 2d). Using a threshold on $Res^{min}$, pixels with large deviations can be rejected from the final SO2 map.

We established that the threshold $Res^{min} = 6 \times 10^{-3}$ can effectively discriminate between reliable and non-reliable $SO_2$ estimations. It should be noted that the $Res^{min}$ threshold depends on the choice of wavelengths in regards to the model spectra and should be computed again for a different set of optical wavelengths. We determined that the number of wavelengths required for the variance method and the $SO_2$ evaluation was at least 18 wavelengths. This large number of optical wavelengths increases the acquisition duration and thereby the need for a motion correction algorithm. However, this number is consistent with other studies that performed accurate determination of the $SO_2$ in vitro [30] with 17 wavelengths, *in vivo* [31] with 10 and 28 wavelengths in preclinical and clinical applications, respectively.

With numerical simulation, we verified that the variance method is precise (low standard deviation) and accurate (low mean error). These performances are linked to the pixel rejection inherent to the variance method.

The variance method was shown to achieve expected $SO_2$ estimations for preclinical experiments during an oxygen challenge, while it has been shown elsewhere that conventional linear unmixing methods fail in anoxic conditions[32]. Moreover, the number of selected pixels after thresholding with the variance method was found to be stable throughout the $O_2$ challenge experiments. The median $SO_2$ value ($SO_2^m$) for 3 experiments on 3 different mice was found to be 64.6% in normoxic conditions (breathing air), 69% for hyperoxia and 44% in hypoxic conditions. During the hyperoxia (100% $O_2$) phase of the oxygen challenge, we found that the $SO_2$ only slightly increases. It has been shown using BOLD MRI that, even though the arterial oxygen saturation is expected to be nearly complete in normoxic conditions, breathing 100% $O_2$ does moderately increase the oxygenation in all the renal compartments[33]. However, we could not find the absolute values for the $SO_2$ increase for mouse kidneys from the literature.

With the variance method and for a significant portion of the selected pixels, the $SO_2$ was below 50% which is low for normal in vivo conditions. These pixels are artefacts resulting from the image processing on the binary mask and further investigations will be considered specifically on the binary mask and an automatic ROI segmentation.

The main source of error in $SO_2$ estimation is the respiratory motion. This is evident from the larger spread of $Res^{min}$ values for in-vivo experiments as compared to the post-mortem experiment (Fig. 2d). Furthermore, the directionality and bandwidth of ultrasound waves generated by blood vessels in MSOT depend on the vessel orientation and size. Hence the emitted OA signals are better captured by ultrasound arrays with a large angular aperture and broad bandwidth[34]. The linear transducer used for this work largely limits resolving the vasculature. However, the motion rejection and unmixing algorithms presented in this article can be adapted to any imaging system which allows the acquisition of simultaneous MSOT and USI images and it will be interesting to apply our methods to next-generation systems with improved efficacy for resolving vasculature [35].

The evaluation of the longitudinal stability of the $SO_2$ measurement in the same animal will be interesting to further validate our method. Monitoring the $SO_2$ have been performed in other studies over several hours [23], [26] but not over several days. Our present experimental set-up was not designed for an easy repositioning of the animal after removal. However, a systematic and standardized study to evaluate the longitudinal fluctuation of $SO_2$ measurements in kidneys of heathy animals will be considered in the future and would be of interest for the photoacoustic community.

For the mouse postmortem, we have shown that our method like the conventional least square method give similar $SO_2$ estimations for pixels with high SNR (higher $Res^{min}$ values). However, a simple amplitude thresholding cannot discriminate between correct and ambiguous values, while a thresholding on $Res^{min}$ can achieve it.

## V. CONCLUSIONS

To conclude, we have presented two image processing methods in order to increase the accuracy of $SO_2$ estimation for preclinical imaging. First, motion artefacts were reduced by selecting only OA images acquired during the respiratory pause of the animal. Furthermore, a simple but robust spectral unmixing method has been proposed to effectively reject pixels where the spectrum is dominated by noise.

Our method outperformed the conventional linear unmixing methods in terms of $SO_2$ measurement accuracy as validated by cross-sectional 2D images and by oxygen challenge experiments in healthy mice. The image processing methods is expected to benefit preclinical studies on murine models of human renal pathologies.

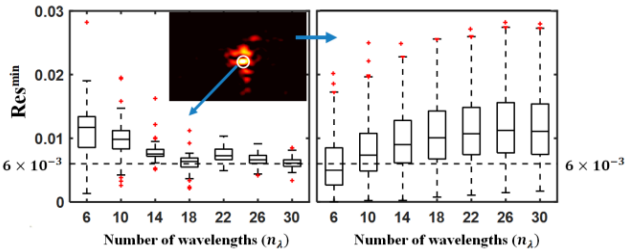

Fig 5. Boxplot of the $Res^{min}$ values as a function of the number of wavelengths used for pixels inside a 300μm inner diameter PTFE tube filled with India ink (left) and outside the tube (right). The OA image of the tube at 780nm and the manual segmentation is shown in inset.

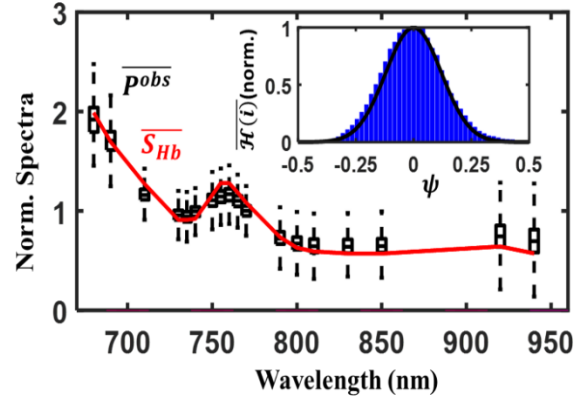

Fig 6. The observed spectra $\overline{P^{obs}}$ and $\overline{S_{Hb}}$, both normalized by their respective mean for 6772 pixels selected by thresholding on Res as shown in Fig 3d. (Inset) The histogram averaged over λ (normalized by its maximum value) for different values of $\psi$. The analytical expression of a normal distribution ($exp((-1/2)(\psi/\sigma)^2)$) with σ=0.11 is also shown (black line).

## APPENDIX

### A. The Variance Method.

The observed raw spectra for MSOT considering linear model is written as

$$\wp^{obs}(\lambda) \propto S_\phi(\lambda) \times [C_{Hb}S_{Hb}(\lambda) + C_{HbO_2}S_{HbO_2}(\lambda) + \Psi(\lambda)]$$
(7)

Where and $S_\phi(\lambda)$ is the incident optical fluence. Using $SO_2 = C_{HbO_2}/(C_{Hb} + C_{HbO_2})$ in Eq.2 we get

$$\frac{\wp^{obs}(\lambda)}{S_\phi(\lambda)} = A \times [(1 - SO_2)S_{Hb}(\lambda) + SO_2\ S_{HbO_2}(\lambda) + \Psi\lambda CHb+CHbO2$$ (8)

Which is same as Equation 2 with $\psi_0 = \Psi(\lambda)/(C_{Hb} + C_{HbO_2})$ and $P^{obs} = \wp^{obs}/S_\phi$. By definition we set $Res(SO_2) = \frac{1}{n_\lambda}\sum_{i=1}^{n_\lambda}|D_m(\lambda_i, SO_2) - \mu Dm(SO2)2$. Hence, $Res(SO_2)$ for the actual value of $SO_2$ has a minimum as a function of $SO_2$ and the value at the minimum ($Res^{min}$) is proportional to the square of the standard deviation (the variance $var$) of $|\Psi(\lambda)|$. It can be shown that

$$Res^{min} = var(|\psi(\lambda)|) = var(\psi(\lambda)) - (E(|\psi(\lambda)|))^2 \quad (9)$$

Where E denotes the expectation of $|\psi(\lambda)|$. If we consider a normal distribution with mean μ=0 and standard deviation σ, then by Equation 9 and $E(|\psi(\lambda)|) = \sigma\sqrt{2/\pi}$, we get $Res^{min} = \sigma^2(1-2/\pi) = 0.36\sigma^2$.

### B. Choice of the number of wavelengths for MSOT.

To determine the number of wavelengths required for an accurate thresholding using the variance method, we used a PTFE tube filled with India Ink which has a featureless absorption spectrum in the range of wavelengths of interest. Fig. 5 illustrates that the $Res^{min}$ value converges toward $6\times10^{-3}$ for $n_\lambda \geq 18$ for pixels inside the tube, while, for pixels outside the tube, Res$^{min}$ mostly stays above this value. This result validates our choice of threshold for Res$^{min}$ and $n_\lambda$=18.

### C. The distribution of the parasitic spectral deviations.

For the experiment of the mouse post-mortem, we calculated the difference between the corrected measured spectra normalized by their mean values ($\overline{P^{obs}}$) and the spectra of Hb ($\overline{P^{model}}(\lambda, SO2 = 0)$) for pixels with Res<6×10$^{-3}$. The histogram of the difference ($\psi(\lambda)$) was calculated for each wavelength ($\mathcal{H}_\lambda(i)$) with i being the equally spaced bins between -1 and +1. Then, the mean of the histogram along λ was calculated ($\overline{\mathcal{H}(i)}$) and shown in Fig 6. A gaussian curve with μ=0 and σ=0.11 is also shown and it fits closely to the histogram of $\psi(\lambda)$. Hence, we can conclude that $\psi(\lambda)$ has a normal distribution and the Res$^{min}$=0.36σ$^2$=4.4×10$^{-3}$ which is close to the experimentally evaluated Res$^{min}$ for the post-mortem experiment (Fig 2d). Spectral coloration would be cumulative and would induce a feature that could not be revealed by the gaussian distribution here. It could indeed be neglected in our study.

## ACKNOWLEDGMENT

The project was supported by Ile-de-France region through a SESAME grant and a Blaise Pascal Chair of Excellence. Imaging was performed at the Imaging Facility (Plateforme d'Imageries du Vivant) of the University of Paris, supported by France Life Imaging (grant ANR-11-INBS-0006) and Infrastructures Biologie-Santé (IBISA). Parts of the technology developed for this study wre supported by SIRIC CARPEM and by Infrastructures Biologie-Santé (IBiSa) grants to B.T. M.S. acknowledges support from the postdoctoral fellowship under the Blaise Pascal International Chairs of Excellence. MP-L acknowledges support from the European Union's Horizon 2020 research and innovation Programme under the Marie Sklodowska-Curie Grant Agreement no.101030046, and the Programme Ramón y Cajal RYC2021-032739-I, funded by MCIN/AEI/10.13039/501100011033 and by the European Union "NextGenerationEU"/PRTR. We are deeply indebted to Prof. Vasilis Ntziachristos for his constant support and advice that he provided during the course of this study as the recipient of the Chair Blaise Pascal from the Ile-de-France region.